# Theoretical prediction of diffusive ionic current through nanopores under salt gradients

Long Ma,[1,2] Zihao Gao,[1,2] Jia Man,[1] Jianyong Li,[1] Guanghua Du,[3] and Yinghua Qiu[1,2,4]*

1. Key Laboratory of High Efficiency and Clean Mechanical Manufacture of Ministry of Education, National Demonstration Center for Experimental Mechanical Engineering Education, School of Mechanical Engineering, Shandong University, Jinan, 250061, China

2. Shenzhen Research Institute of Shandong University, Shenzhen, 518000, China

3. Institute of Modern Physics, Chinese Academy of Sciences, Lanzhou, 730000, China

4. Suzhou Research Institute of Shandong University, Suzhou, 215123, China

*Corresponding author: yinghua.qiu@sdu.edu.cn

**ABSTRACT**

In charged nanopores, ionic diffusion current reflects the ionic selectivity and ionic permeability of nanopores which determines the performance of osmotic energy conversion, i.e. the output power and efficiency. Here, theoretical predictions of the diffusive currents through cation-selective nanopores have been developed based on the investigation of diffusive ionic transport under salt gradients with simulations. The ionic diffusion current *I* satisfies a reciprocal relationship with the pore length $I \propto \alpha/L$ ($\alpha$ is a constant) in long nanopores. $\alpha$ is determined by the cross-sectional areas of diffusion paths for anions and cations inside nanopores which can be described with a quadratic power of the diameter, and the superposition of a quadratic power and a first power of the diameter, respectively. By using effective concentration gradients instead of nominal ones, the deviation caused by the concentration polarization can be effectively avoided in the prediction of ionic diffusion current. With developed equations of effective concentration difference and ionic diffusion current, the diffusion current across nanopores can be well predicted in cases of nanopores longer than 100 nm and without overlapping of electric double layers. Our results can provide a convenient way for the quantitative prediction of ionic diffusion currents under salt gradients.

**Keywords:** Ionic diffusion; Nanopores; Osmotic energy conversion; Electric double layers.

**I. INTRODUCTION**

Osmotic energy provides a kind of renewable and green energy source, which may alleviate the energy crisis faced by human society.[1-3] With charged porous membranes, selective diffusion of counterions under concentration gradients happens due to the electrostatic interaction between surface charges on pore walls and free ions in the solution, which induces a considerable electrical potential across the membrane.[4, 5] Then, osmotic energy conversion (OEC) is achieved which converts the chemical energy in the system to the electrical energy.[3, 6]

As important parameters to evaluate the OEC performance, the output power and OEC efficiency are closely related to the ionic selectivity and permeability of nanopores.[4, 5, 7] A clear understanding of the ionic diffusive transport is essential because the induced ionic diffusion current reflects the ionic flux and nanopore selectivity. Fick's law describes the diffusion transport behavior of ions under concentration gradients in bulk solutions.[8-10] In neutral nanopores, without the influences of surface charges, the ionic diffusion can be described by Fick's law, which is determined by the concentration gradient and ionic diffusion coefficients.[11] While in charged nanopores, the diffusive transport of ions becomes more complicated.[12] A large number of counterions accumulate in electric double layers (EDLs) near the charged pore wall.[13, 14] The ionic diffusive transport in the EDLs contributes a large portion of the diffusion current, which cannot be predicted by Fick's law.[11] In addition, due to the appearance of concentration polarization across the porous membrane, weakened salt gradients than the nominal ones lead to lower diffusive currents through nanopores which decrease the OEC performance.[15-17]

Inside charged nanopores, the ionic diffusive transport can be determined by the pore dimensions,[16] surface charges,[7, 18, 19] and surface slip length,[20, 21] which affect the ionic selectivity and permeability of nanopores. With concentration differences across nanopores, the pore length determines the concentration gradient. During the diffusive transport of ions inside nanopores, the pore diameter controls the diffusion flux. Surface charges induce the diffusive transport of counterions in EDLs regions, i.e. surface diffusion, which leads to a high selectivity to counterions in long nanopores.[16] In thin nanopores, charged external surfaces on the low-concentration side can provide effective promotion in the diffusive transport of counterions by enlarging the effective diffusion area.[7, 19, 20] Directional ionic diffusion induces the fluid movement due to ionic hydration, i.e. diffusio-osmotic flow, which enhances both diffusive transport of cations and anions.[11] Under confinement, diffusion-osmotic flow can be improved by slippery surfaces because of the reduced flow friction.[20]

Siwy et al.[22] investigated the ionic diffusion current across conical polymer nanopores under salt gradients. Due to the asymmetric pore shape i.e. different sizes at both pore ends, the direction of the concentration gradient affected the diffusion current. A concentration gradient from the base to the tip induced a greater current, which was due to the weaker ionic selectivity of the pore tip in solutions of a higher concentration. The diffusion current was semiquantitatively described with the Smoluchowski Nernst-Planck equations. With boron nitride nanotubes, Bocquet et al.[23] studied the OEC performance under concentration gradients of KCl. They developed a theoretical formula to characterize the net diffusion current in nanopores with the micrometer length and diameters much larger than the EDLs thickness under concentration gradients. With the applied salinity ratio, the diffusion current can be predicted by the diffusio-osmotic mobility and ionic concentration. Due to the similar diffusion coefficients for $K^+$ and $Cl^-$ ions, coions were not considered in their theoretical predictions.

To provide a quantitative description of ionic diffusion current, the finite element method (FEM) was employed to investigate the ionic diffusive transport in nanopores under various conditions. For short nanopores, due to the appearance of concentration polarization,[15, 16] the effective concentration difference at both ends of the nanopore is much smaller than the nominal one. The effective concentration difference at both pore openings is found to provide the real driving force for ionic diffusion. Based on the simulations, we raised theoretical predictions of effective concentration differences for cations and anions across the nanopore, as well as the total effective ion concentrations on both high- and low-concentration sides. Considering the respective transport passageways of cations and anions in charged nanopores, ionic diffusive transport in both the central-pore region and the EDLs region near charged surfaces are included to describe the ionic diffusion current.[19] Using the effective concentration differences, equations were developed to describe ionic diffusion currents, which can be applied under various conditions.

**II. METHODOLOGY**

COMSOL Multiphysics was used to solve 3D simulation models, in which Poisson-Nernst-Planck (PNP) and Navier-Stokes (NS) equations (Eqs. 1-4) are coupled to describe the transport behaviors of ions and fluids in the reservoirs and the nanopore.[24-26]

$$\varepsilon \nabla^2 \varphi = -\sum_{i=1}^{N} z_i F C_i \quad (1)$$

$$\nabla \cdot \mathbf{J}_i = \nabla \cdot \left( C_i \mathbf{u} - D_i \nabla C_i - \frac{F z_i C_i D_i}{RT} \nabla \varphi \right) = 0 \quad (2)$$

$$\mu \nabla^2 \mathbf{u} - \nabla p - \sum_{i=1}^{N} (z_i F C_i) \nabla \varphi = 0 \quad (3)$$

$$\nabla \cdot \mathbf{u} = 0 \quad (4)$$

where $\nabla$, $\varphi$, $R$, $T$, $F$, and $p$ are the gradient operator, electrical potential, gas constant, temperature, Faraday's constant, and pressure, respectively. $\varepsilon$, $\mu$, and $\mathbf{u}$ represent the dielectric constant, viscosity, and velocity of solutions. $z_i$, $\mathbf{J}_i$, $C_i$, and $D_i$ are the valence, ionic flux, concentration, and diffusion coefficient of species $i$, which denotes the cation or anion. $N$ is the number of ion species in aqueous solutions.

As shown in Figure 1, 3D simulation models were constructed by connecting two reservoirs through a nanopore.[4,5] Both reservoirs were 5 μm in radius and height. The length of the nanopore ($L$) was changed from 1 nm to 10 μm. The pore diameter ($d$) was set from 3 to 30 nm, with a default pore size of 10 nm which can be matched with nanofluidic experiments.[27-31] The inner pore wall carried surface charges.[16] The surface charge density ($\sigma$) varied from 0 to −0.16 C/m$^2$, with a default value of −0.08 C/m$^2$.[7, 15, 32]

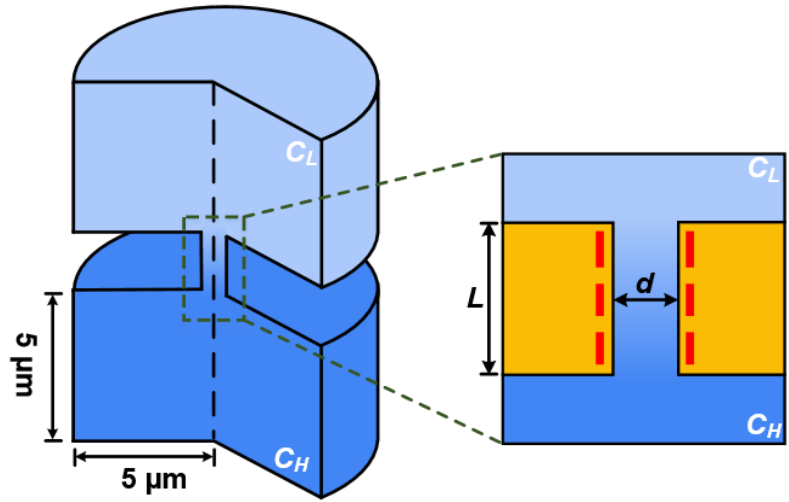

Figure 1 Schematic illustration of simulations.

In this work, three ionic solutions were considered, i.e. LiCl, NaCl, and KCl. A series of salt gradients were applied across the nanopore, with the low salt concentration ($C_L$) maintained at 10 mM, and the high concentration ($C_H$) varying from 100 mM to 1000 mM. NaCl solution under a concentration gradient of 50-fold is the default condition, which represents the concentration gradient of the river water and seawater at the natural estuaries.[33, 34] Diffusion coefficients of K$^+$, Na$^+$, Li$^+$, and Cl$^-$ ions were set as 1.96×10$^{-9}$, 1.33×10$^{-9}$, 1.03×10$^{-9}$ and 2.03×10$^{-9}$ m$^2$s$^{-1}$, respectively.[35] The temperature and the relative permittivity of water were 298 K and 80. Detailed boundary conditions are listed in Table I. The meshing strategy is provided in Figure S1. [7, 20, 36]

The diffusion current $I$ was calculated by the integration of the total ion flux ($\mathbf{J}_i$) over the reservoir boundary with Eq. 5.[7, 37, 38]

$$I = \int_S F\left(\sum_i^2 z_i \mathbf{J}_i\right) \cdot \mathbf{n} \, dS \tag{5}$$

where **n** and $S$ denote the unit normal vector, and the reservoir boundary, respectively.

Table I Boundary conditions used in numerical modeling. Coupled Poisson-Nernst-Planck and Navier-Stokes equations were solved with COMSOL Multiphysics.

| Scheme | Surface | Poisson | Nernst-Planck | Navier-Stokes |
|---|---|---|---|---|
| | AB | constant potential (Ground) $\varphi=0$ | constant concentration $C_i=C_L$ | constant pressure $p=0$ no viscous stress $\mathbf{n}\cdot[\mu(\nabla\mathbf{u}+(\nabla\mathbf{u})^T)]=0$ |
| | BC, FG | no charge $-\mathbf{n}\cdot(\varepsilon\nabla\varphi)=0$ | no flux $\mathbf{n}\cdot\mathbf{N}_i=0$ | no slip |
| | HG | constant potential $\varphi=0$ | constant concentration $C_i=C_H$ | constant pressure $p=0$ no viscous stress $\mathbf{n}\cdot[\mu(\nabla\mathbf{u}+(\nabla\mathbf{u})^T)]=0$ |
| | AH | axial symmetry | axial symmetry | axial symmetry |
| | DC, EF | no charge $-\mathbf{n}\cdot(\varepsilon\nabla\varphi)=0$ | no flux $\mathbf{n}\cdot\mathbf{N}_i=0$ | no slip |
| | DE | $-\mathbf{n}\cdot(\varepsilon\nabla\varphi)=\sigma$ | no flux $\mathbf{n}\cdot\mathbf{N}_i=0$ | no slip $\mathbf{u}=0$ |

**III. RESULT AND DISCUSSIONS**

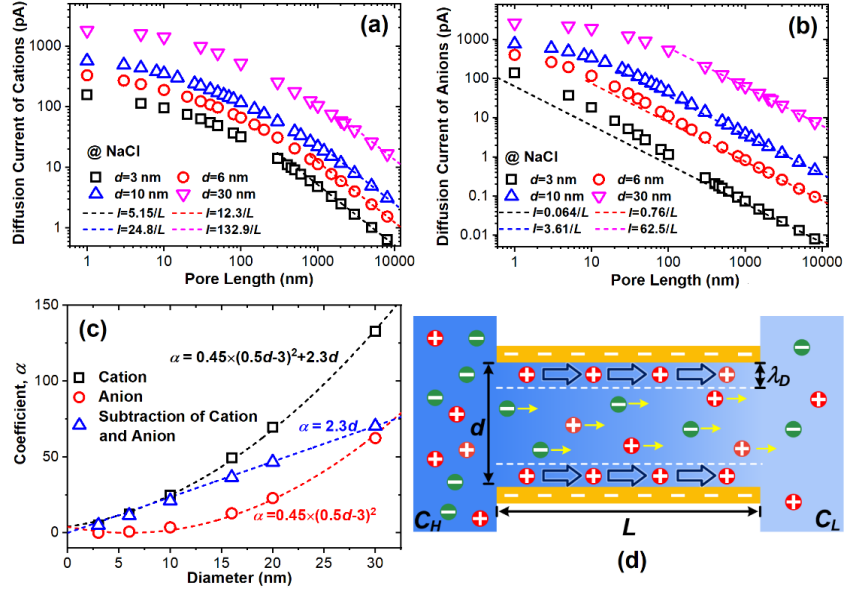

Figure 2 Diffusion current through differently wide nanopores with various pore lengths in NaCl solution. (a) Diffusion current of cations, (b) diffusion current of anions, and (c) coefficient $\alpha$ in the fitting equation $\alpha/L$ at various pore diameters. (d) Illustrations of the ionic transport under salt concentration gradients. Ionic diffusion inside the EDLs near charged pore walls and the center of the nanopore are shown by the dark blue and yellow arrows, respectively. The concentration gradient was set to 500: 10 mM NaCl. $C_H$ and $C_L$ denote the high and low salt concentrations. $\lambda_D$ denotes the thickness of EDLs, which is approximately equal to the Debye length.

Under salinity gradients, both cations and anions have diffusive transport through nanopores from the high-concentration side to the low-concentration side, generating the diffusion current. In negatively charged nanopores, due to the formation of EDLs, the diffusive transport of cations happens in the EDLs region and the central region of the nanopore, which results in surface conductance and bulk conductance, respectively.[19, 39] Because of the electrostatic repulsion from surface charges, anions diffuse across the nanopore only through the pore center. Taking advantage of the simulation method, the diffusion current contributed by cations and anions can be obtained separately. Due to the negative valence of anions, the absolute values of the diffusion current caused by Cl$^-$ ions were used for data analysis.

In the simulations, the nanopore serves as a passageway for ion diffusion. The length and diameter of the nanopore determine the distance and cross-sectional area for the directional ionic diffusion, respectively.[19] In this study, the diffusion of cations and anions was investigated in nanopores with various lengths and diameters under the natural salt gradient of 500:10 mM NaCl. The nanopore length was changed from 1 nm to 10 μm, and the diameter was varied from 3 to 30 nm. From Figures 2a and 2b, when the same concentration difference exists across nanopores with different lengths, according to Fick's law,[10] the driving force for ion diffusion increases as the nanopore length decreases, leading to an increased flux of both cations and anions. The charged nanopore has the selectivity to counterions, which is positively and negatively correlated with the pore length and the diameter, respectively.[40] For long nanopores with a diameter below 10 nm, the diffusion current of cations is greater than that of anions, although $Cl^-$ ions have a larger diffusion coefficient. While in short nanopores the electrostatic repulsion between negative surface charges and $Cl^-$ ions is weak, which results in a larger diffusion current of $Cl^-$ ions than $Na^+$ ions.[7]

The increase in the nanopore diameter enlarges the cross-sectional area for ion diffusion, which reduces the impact of charged surfaces on ion transport.[12] The nanopore exhibits a weaker ionic selectivity to counterions. In wider nanopores, both diffusive transport of cations and anions increases in the central region of the nanopore. Due to the decreased proportion of cation diffusion flux in the EDLs region, $Cl^-$ ions with a larger diffusion coefficient exhibit a higher diffusion current than $Na^+$ ions. In the nanopore with a diameter larger than 6 nm and a length less than 100 nm, the cation transfer number $t_+$ is less than 0.5, i.e. the nanopore is selective to anions (Figure S2).[26]

Ion diffusion flux is proportional to the concentration gradient, which is determined by the ratio of the concentration difference to the nanopore length.[10] In the simulation, the concentration difference on both sides of the nanopore remains constant. From Figure 2, for nanopores longer than ~1 μm, the diffusion current of cations has a reciprocal relationship with the pore length, $I \sim \alpha/L$.[10] Similarly, the reciprocal relationship between anion current and the pore length can appear in nanopores longer than ~300 nm. However, when the pore length becomes shorter, the diffusion current of cations and anions no longer follows the $\alpha/L$ relationship. This is mainly due to the huge concentration gradient across the ultrashort nanopores, which can induce significant ion diffusion through the nanopore. The high ionic diffusive flux dramatically decreases the ion concentration at the pore entrance on the high-concentration side and increases the ion concentration at the exit on the low-concentration side, which results in a significant reduction in the effective concentration difference across the nanopore.[15, 16] The reduced salt difference lowers the driving force for ion diffusion, leading to a smaller diffusion flux. Due to the weak ion supplement from the high-concentration bulk solution to the pore entrance and the slow ion transport from the pore exit to the low-concentration bulk solution, obvious ion depletion and enrichment appear on the high- and low-concentration sides of the short nanopores, respectively, which is similar to the ion concentration polarization phenomenon observed in charged nanopores under high voltages.[41]

While for long nanopores, the relatively small concentration gradient results in a lower diffusion flux inside the nanopore, which cannot form significant ionic depletion and enrichment at both ends of the nanopore. From the theoretical prediction with $\alpha/L$ to the diffusion current of anions and cations in nanopores with different diameters, the relationship between the coefficient $\alpha$ and the pore diameter is obtained, as shown in Figure 2c. For the diffusion current of $Cl^-$ ions, the coefficient $\alpha$ exhibits a quadratic relationship with the diameter. While for the diffusion current of $Na^+$ ions, the relationship between $\alpha$ and the pore diameter has two components that are superimposed with a power relationship based on the quadratic relationship with the diameter. Considering the diffusion paths of cations and anions in the nanopores,[19] the relationship between the coefficient $\alpha$ and the pore diameter can be quantitatively explained.

Figure 2d illustrates the schematic of diffusive transport for cations and anions in a long nanopore. The EDLs, with a thickness approximately equal to the Debye length ($\lambda_D$),[14] near the charged pore walls provide a fast diffusion passageway for cations.[7, 19] Away from the pore surfaces, both cations and anions can undergo diffusive transport in the central region of the nanopore. Under concentration gradients, inhomogeneous concentration distributions appear in the nanopore varying from the high concentration to the low one. Since the $\lambda_D$ depends on the solution concentration, it decreases with the increase of the solution concentration. Here, we fix a specific $\lambda_D$ for each case because of the small proportion of the EDLs region inside the nanopore. The surface area of the EDLs region ($S_{EDLs}$) can be roughly estimated by $S_{EDLs}=\pi D\lambda_D$.[19] Then, the cross-sectional area of the central region ($S_{center}$) can be calculated with $S_{center}=\pi(D/2-\lambda_D)^2$.[19] Since the inner surface of the nanopore carries negative charges, the diffusive transport of Cl$^-$ ions mainly happens in the central region of the nanopore. Na$^+$ ions, as the counterions, can simultaneously diffuse in the EDLs region and the central region of the nanopore. Inside the nanopore, the variation of the cross-sectional area for the diffusive transport of anions and cations with the pore diameter is exactly consistent with the relationship between the coefficient $\alpha$ and the pore diameter in Figure 2c.

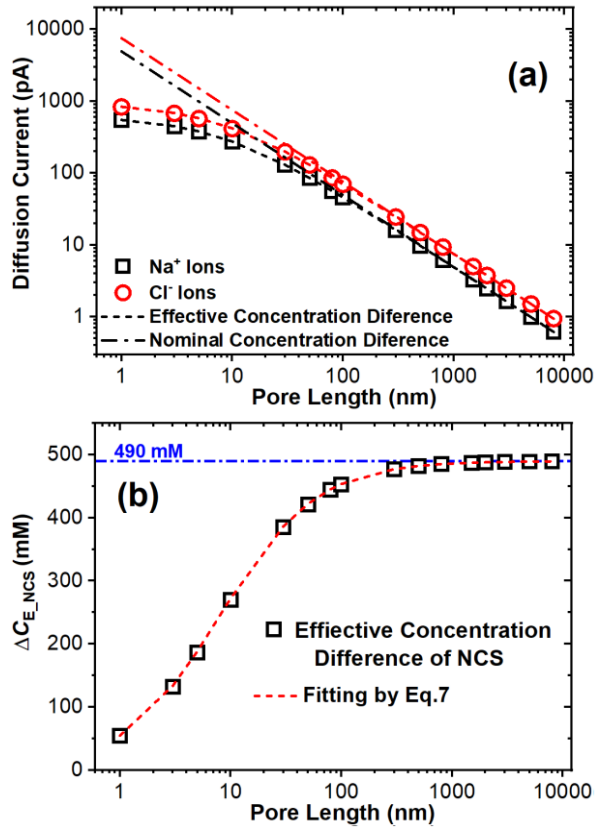

Figure 3 Diffusion current and effective concentration difference across the pore as a function of the pore length in nanopores with no charged surfaces (NCS). (a) Diffusion current of Na$^+$ and Cl$^-$ ions. The dash-dotted lines and short-dashed lines represent the nominal and effective concentration differences used in Eq. 6, respectively. (b) Effective concentration difference across nanopores. The red-short-dashed line is the theoretical fitting with Eq. 7. The nanopore diameter is 10 nm. The blue-dash-dotted line represents the nominal concentration difference between both ends of the nanopore, i.e. 490 mM.

The quantitative description of the ionic diffusion current through nanopores under salt gradients is crucial for elucidating the mechanism of ion transport under confinement, which may find important application in high-performance osmotic energy conversion,[3] seawater desalination,[42] and other fields.[43] In aqueous solutions, ion transport can be described by the Nernst-Planck equation (Eq. 2). Under the concentration gradients, the ion transport in nanopores is primarily governed by diffusion, while contributions from migration and convection can be neglected.[12] Then, the ionic diffusion current can be estimated by Eq. 6, which does not account for the effect of surface charges. Figure 3 shows the diffusion currents of cations and anions in neutral nanopores of different lengths, with the nominal concentration difference Eq. 6 (dash-dotted lines) can predict the ionic diffusion current $I_i$ ($i$ is Na$^+$ and Cl$^-$) in nanopores longer than 50 nm accurately.[10]

$$I_i \approx \pi(\frac{d}{2})^2 FD_i \frac{\Delta C}{L} \quad (6)$$

where $\Delta C$ is the nominal concentration difference between the high- and low-concentration sides.

$$\Delta C_{E\_NCS} = (C_H - C_L) \times \frac{L}{0.9d + L} \quad (7)$$

where $\Delta C_{E\_NCS}$ is the effective concentration difference between the high- and low-concentration side in the neutral nanopore.

$$I_{NCS} \approx \pi(\frac{d}{2})^2 FD_i \frac{\Delta C_{E\_NCS}}{L} = \pi(\frac{d}{2})^2 FD_i \frac{(C_H - C_L)}{0.9d + L} \quad (8)$$

where $I_{NCS}$ is the diffusion currents of cations and anions in the neutral nanopore with the effective concentration difference between the high- and low-concentration side in the nanopore.

In Figure 3a, for neutral nanopores shorter than 50 nm, the relationship between ionic diffusion current and the pore length no longer follows $I \propto α/L$, which is due to the obvious concentration polarization across the nanopore.[16] Taking advantage of simulations, the effective concentration can be extracted at both ends of the nanopore, i.e. the averaged concentrations within the cross-sections at the entrance and exit. As shown in Figure 3b, for nanopores longer than ~500 nm, the effective concentration difference at both pore ends is equal to the nominal concentration difference i.e. 490 mM. However, as the nanopore length shortens, the deviation between the nominal and effective concentration differences becomes more significant which can reach ~440 mM for a pore length of 1 nm.[16]

Equation 7 is proposed to fit the effective concentration difference at both ends of neutral nanopores with various lengths. With the application of the effective concentration difference in Eq. 6, the diffusion current in neutral nanopores can be perfectly predicted with Eq. 8, which indicates that the effective concentration difference is the better choice for diffusion current prediction than the nominal one. The derivation of Equations (6) to (8) was added in the Supplementary Material.

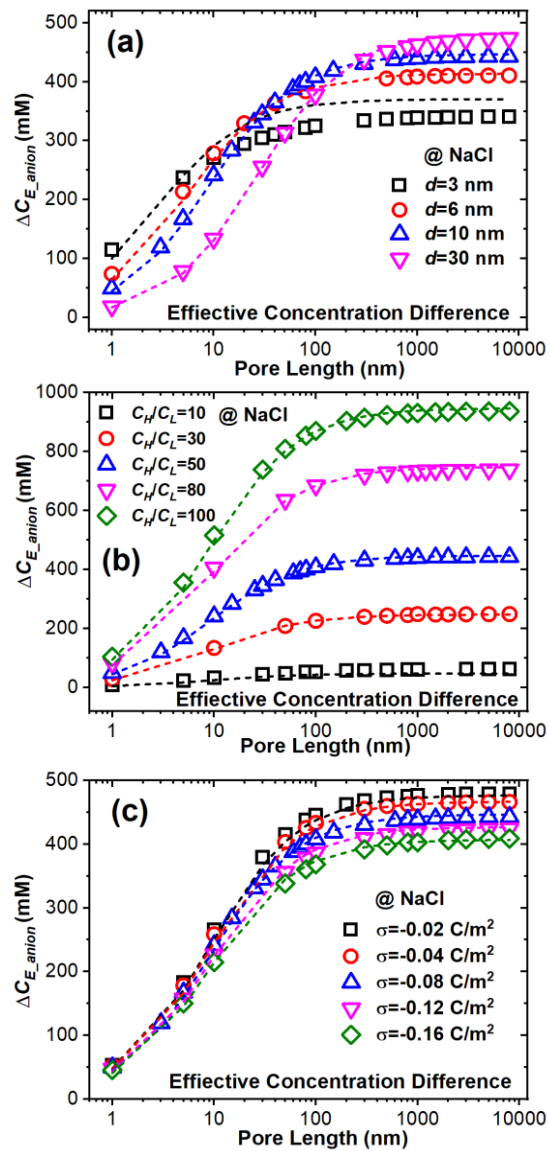

Figure 4 Effective concentration difference of anions and theoretical fitting with Eq. 9 under different conditions in NaCl solution. (a) Diameter, (b) nominal concentration gradient, and (c) surface charge density.

With charged inner surfaces, the diffusive transport of ions in the nanopore becomes complicated due to the electrostatic interaction between the surface charges and free ions which is more pronounced in nanopores with smaller diameters.[12] From Figure S3, Equation 6 exhibits significant deviations in fitting the diffusion current of $Na^+$ ions and $Cl^-$ ions, which may be only suitable for the prediction of ion transport in neutral nanopores since the effect of surface charges is not considered. In nanopores with charged inner surfaces, $Na^+$ ions, as counterions, exhibit surface diffusion in the EDLs regions along charged walls, which serves as the main contribution to the diffusion current of $Na^+$ ions.[19] $Cl^-$ ions mainly transport in the central region of the nanopore where the thickness of the EDLs on the low-concentration side ($\lambda_D \approx 3$ nm) cannot be neglected in narrow nanopores. Without consideration of the EDLs' thickness, Eq. 6 overestimates the cross-sectional area for the diffusive transport of $Cl^-$ ions.

With charged inner surfaces nanopores exhibit selectivity to counterions, which affects the effective ion concentration at the nanopore entrance and exit.[19] From Figure 4a, for nanopores shorter than ~100 nm, a smaller diameter can result in a larger effective concentration difference at both pore ends due to the more restricted diffusive transport of $Cl^-$ ions under the higher confinement. As the pore length increases, the $Cl^-$ ion concentration inside the nanopore decreases because of the increased cation selectivity.[40] The effective concentration of $Cl^-$ ions at the entrance of narrow nanopores is significantly lower than that of wide ones (Figure S4a), which induces a smaller effective concentration difference at both pore ends. As the bulk concentration gradient increases, the effective concentration difference at both pores ends enhances (Figure 4b). Under a stronger surface charge density, the promoted electrostatic repulsion between surface charges and $Cl^-$ ions induces a smaller $Cl^-$ ion concentration at the pore entrance (Figure S4c). The effective concentration difference at both pores ends becomes smaller as the surface charge density increases (Figure 4c). Eq. 9 is raised with the consideration of the surface charge density, nanopore dimensions, and the effective concentration gradient across the nanopore. As shown in Figure 4, Eq. 9 can provide a perfect prediction of the effective concentration difference of Cl⁻ ions across the nanopore under various conditions, except the only case with overlapped EDLs ($d$=3 nm).

$$\Delta C_{E\_anion} = C_{EH\_anion} - C_{EL\_anion} = [(C_H - C_L) - 184 e^{(-d/6.33)} - 500|\sigma| + 35] \times \frac{L}{(0.9d) + L} \quad (9)$$

in which $\Delta C_{E\_anion}$ is the effective concentration difference of anions across the nanopore. $C_{EH\_anion}$ and $C_{EL\_anion}$ are the effective concentration of anions at the high- and low-concentration sides, respectively.

$$I_{anion} \approx \frac{1}{2}\pi[(\frac{d}{2} - \lambda_{DH})^2 + (\frac{d}{2} - \lambda_{DL})^2]FD_{anion}\frac{\Delta C_{E\_anion}}{L} \quad (10)$$

where $I_{anion}$ is the diffusion current contributed by anions. $\lambda_{DH}$ and $\lambda_{DL}$ are the thickness of the EDLs on the high- and low-concentration sides. $D_{anion}$ is the diffusion coefficient of anions.

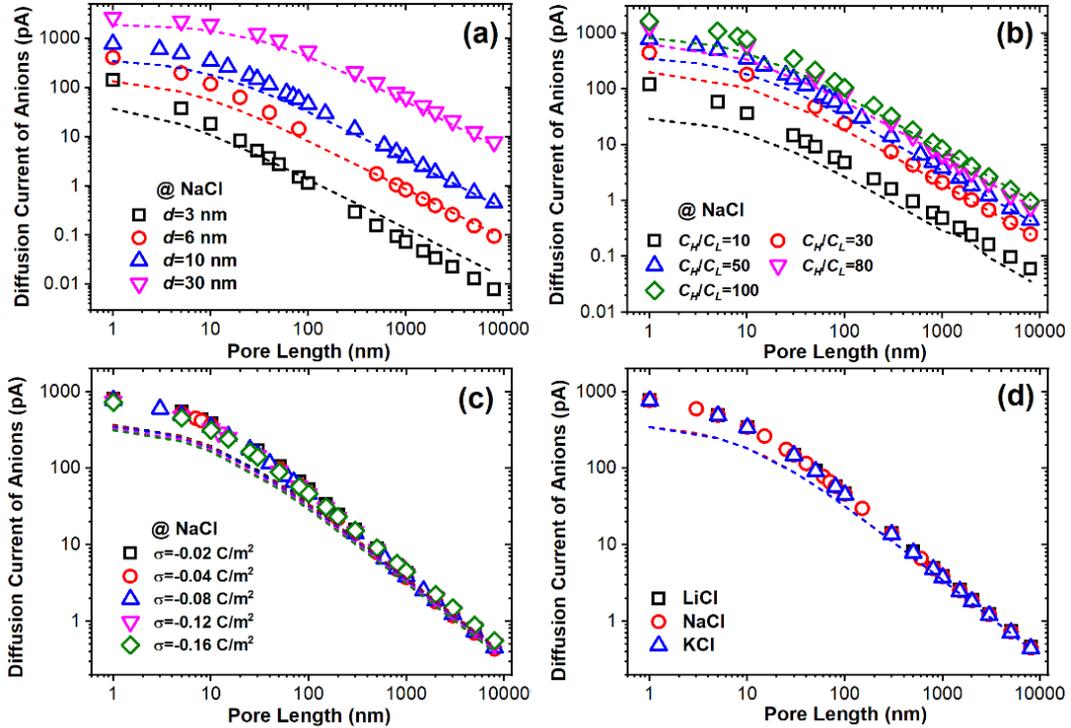

Figure 5 Diffusion current of anions with various pore lengths under different conditions. (a) Diameter, (b) nominal concentration gradient, (c) surface charge

density, and (d) salt types. Short-dashed lines are the theoretical fitting with Eq. 10.

Figure 5 illustrates the diffusion current of anions through nanopores with different lengths under various simulation conditions. From Figure 5a, under a constant pore length, the high- and low-concentration solutions across nanopores induce almost unchanged $\lambda_D$ in nanopores with different diameters. While the central region of the nanopore expands with increased pore diameter where more $Cl^-$ ions can diffuse through, leading to an enhanced diffusion current of $Cl^-$ ions.[19] The strength of ion diffusion depends on the concentration difference across the nanopore. With a higher concentration difference, the diffusion current of $Cl^-$ ions increases (Figure 5b). Although a higher surface charge density leads to stronger electrostatic repulsion between charged surfaces and $Cl^-$ ions, the diffusion current of $Cl^-$ ions remains unchanged due to the screening to surface charges by counterions (Figure 5c).[14] In LiCl, NaCl, and KCl solutions, different cations do not affect the diffusion current of $Cl^-$ ions. Equation 10 is derived by considering the effective concentration difference of $Cl^-$ ions between the entrance and exit of the nanopore, and the cross-sectional area for the diffusion of $Cl^-$ ions. From Figure 5, Eq. 10 provides a good theoretical prediction of the diffusion current of anions $I_{anion}$ in nanopores under various simulation conditions.

Equation 6 can accurately describe the ionic diffusion current in the central region of the nanopore. Since without consideration of surface charges, Eq.6 fails to capture the diffusive transport of $Na^+$ ions in the EDLs, which induces a lower theoretical prediction of $Na^+$ ionic diffusion current than the simulation results (Figure S3). With boron nitride nanotubes of ~1 μm in length and 30–80 nm in diameter, Bocquet et al.[23] detected the diffusion current through these nanotubes under different salt concentration gradients. Eq. 11 was developed by them to describe the net diffusion current of ions, which is contributed by the counterion transport in EDLs near charged surfaces. In their research with KCl solutions, the net diffusion current generated in the central region of the nanopore is almost zero, due to the similar diffusion coefficient of K$^+$ ions and Cl$^-$ ions. In our simulations, since the diffusion coefficient of Cl$^-$ ions is larger than that of Na$^+$ and Li$^+$ ions, a larger diffusion flux of Cl$^-$ ions appears in the central region of the nanopore than that of Na$^+$ and Li$^+$ ions. In cases with NaCl and LiCl solutions, the net diffusion current $I_{net}$ in the central region of the nanopore is negative, which cannot be predicted by Eq. 11 accurately.

$$I_{net} \approx \frac{\pi d \sigma}{L} \frac{k_B T}{\eta \lambda_B} \Delta \log(\frac{C_H}{C_L}) \qquad (11)$$

where $k_B$, and $\eta$ are the Boltzmann constant, and water viscosity, respectively. $\lambda_B$ is the Bjerrum length, which is equal to 0.7 nm in water.[23]

$$I_{cation} \approx \frac{\pi d \sigma}{L} \frac{k_B T}{\eta \lambda_B} \Delta \log(C_{EH\_total}/C_{EL\_total}) + \pi(\frac{d}{2})^2 FD_{cation} \frac{\Delta C_{E\_cation}}{L} \qquad (12)$$

where $I_{cation}$ is the diffusion current contributed by cations. $C_{EH\_total}$, and $C_{EL\_total}$ are the effective total concentrations at both pore orifices on the high- and low-concentration sides. $\Delta C_{E\_cation}$ is the effective concentration difference of Na$^+$ ions across the nanopore. $D_{cation}$ is the diffusion coefficient of cations.

$$\Delta C_{E\_cation} = C_{EH\_cation} - C_{EL\_cation} = [(C_H - C_L) + 77e^{(-d/4.43)} + 75|\sigma| - 8] \times \frac{L}{(0.9d) + L} \qquad (13)$$

where $C_{EH\_Na}$, and $C_{EL\_Na}$ are the effective concentration of Na$^+$ ions at the high- and low-concentration sides, respectively.

$$C_{EH\_total} = C_{EH\_cation} + C_{EH\_anion} = \frac{(2C_H + 375|\sigma|) \times L^{0.5}}{[0.8(d/10)]^{0.5} + L^{0.5}} \qquad (14)$$

where the $C_{EH\_total}$ is the total effective concentration on the high concentration side.

$$\begin{aligned} C_{EL\_total} &= C_{EH\_total} - \Delta C_{E\_cation} - \Delta C_{E\_anion} \\ &= C_{EH\_cation} + C_{EH\_anion} - (C_{EH\_cation} - C_{EL\_cation}) - (C_{EH\_anion} - C_{EL\_anion}) \\ &= C_{EL\_cation} + C_{EL\_anion} \end{aligned} \qquad (15)$$

Eq. 12 is obtained with the combination of Eqs. 6 and 11, which theoretically describe the diffusion currents of counterions in EDLs near charged inner surface and in the central region of the nanopore, respectively. Please note when the thickness of EDLs is much smaller than the radius of the nanopore with a microscale length, Eq. 11 provides a better theoretical prediction of the diffusion current, because the effective concentration difference across the nanopore approximates the nominal one. While for short nanopores, due to the large deviation between the effective and nominal concentration differences, the application of the effective concentration difference $\Delta C_{E\_cation}$ in Eq. 12 is essential, which is calculated using Eq. 13.

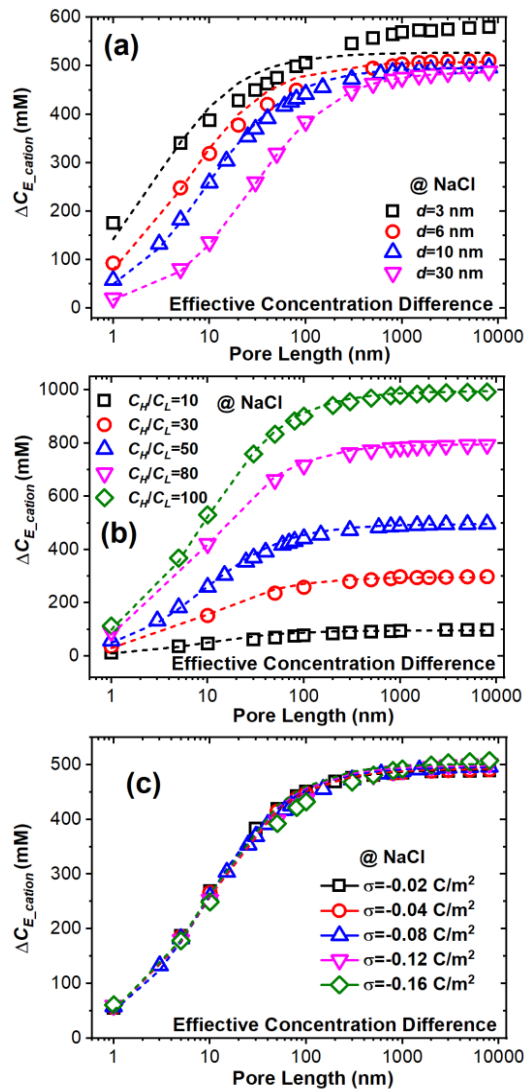

Figure 6 Effective concentration difference of cations and theoretical fitting with Eq. 13 under different conditions in NaCl solution. (a) Diameter, (b) nominal concentration gradient, and (c) surface charge density.

From Figure 6, the effective concentration difference of Na$^+$ ions across the nanopore shares a similar trend to that of Cl$^-$ ions which increases with the pore length. As the diameter decreases, the proportion of the EDLs region inside the nanopore increases, which induces a higher cation selectivity. In narrower nanopores, higher effective concentrations of Na$^+$ ions appear at the entrance and exit (Figure S5a). However, the effective concentration difference of Na$^+$ ions at both ends of long nanopores lowers with the diameter which is opposite to the trend of Cl$^-$ ions (Figure 6a). Due to the dependence of the effective concentration difference on the concentrations at both pore ends, the effective concentration difference of Na$^+$ ion enhances with the applied concentration gradient (Figure 6b). For differently charged nanopores, the effective concentration difference of Na$^+$ ions exhibits independence on the surface charge density. This is attributed to the similar modulation of the surface charge density on the effective concentration of Na$^+$ ions at the nanopore entrance and exit (Figure S5c). In Figure 6, $\Delta C_{E\_cation}$ can be well fitted by Eq. 13 except for the case with overlapped EDLs at $d$ = 3 nm. The total effective concentrations including both Cl$^-$ and Na$^+$ ions at the high-concentration side $C_{EH\_total}$ and the low-concentration side $C_{EL\_total}$ under different conditions can be well described by Eqs. 14 and 15, respectively (Figure S6).

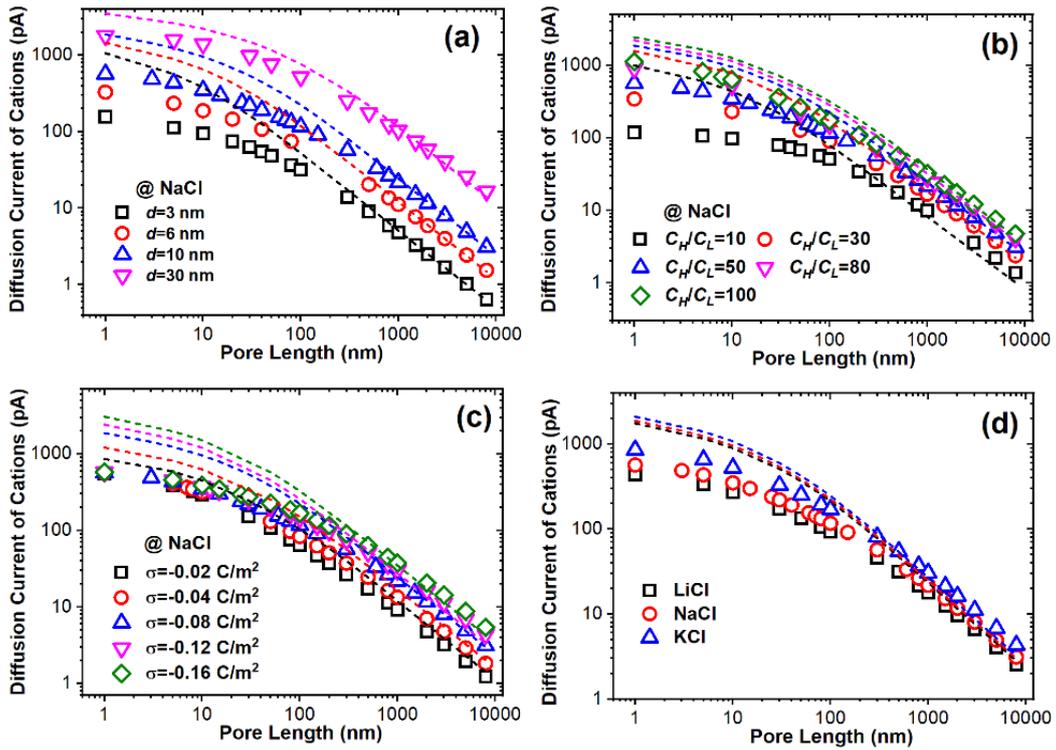

Figure 7 Diffusion current of cations through nanopores with various lengths under different conditions. (a) Diameter, (b) nominal concentration gradient, (c) surface charge density, and (d) salt types.

Figure 7 presents the variation of the counterion diffusion current with pore length under different conditions in NaCl solution. For nanopores longer than 100 nm, Eq. 12 can effectively describe the diffusion current of cations $I_{cation}$ under various pore diameters, applied salt gradients, and surface charge density, which also applies to simulation cases with LiCl and KCl solutions. Compared with the equation of Bocquet et al.[23] (Eq. 11) which has harsh requirements on the solution concentration and pore diameter, the application of Eq. 12 only requires non-overlapped EDLs on the low-concentration side.

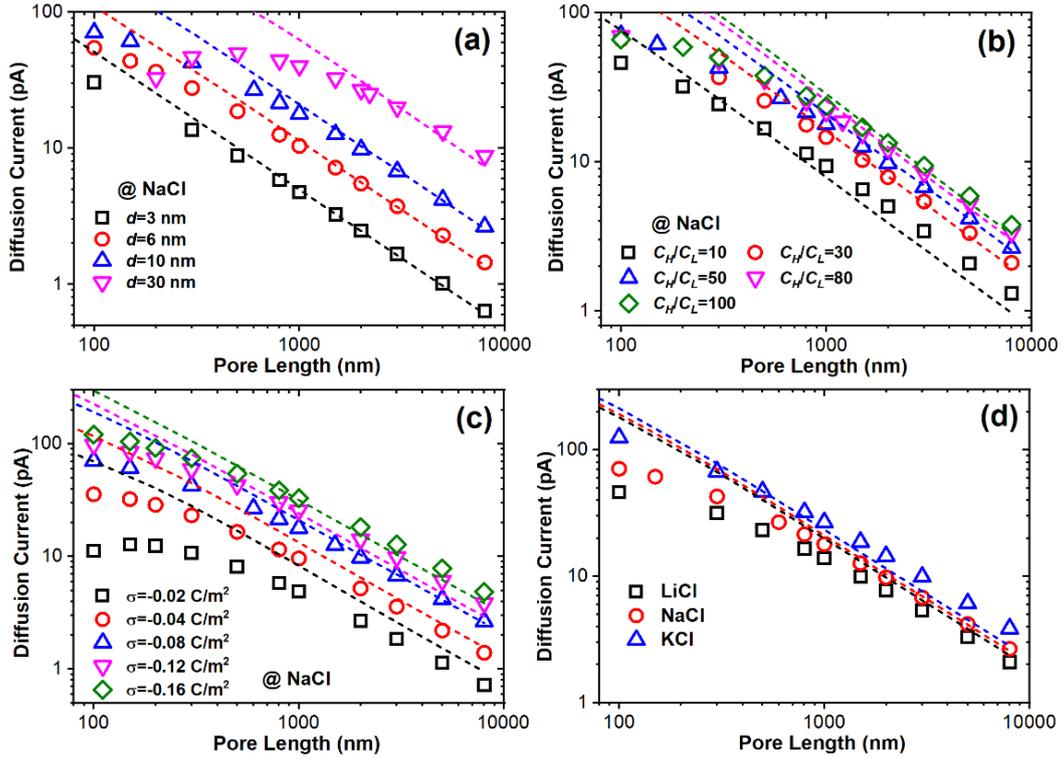

Figure 8 Net diffusion current through nanopores with various lengths under different conditions. (a) Diameter, (b) nominal concentration gradient, (c) surface charge density, and (d) salt types. Short-dashed lines are theoretically fitting with Eq.14.

$$\begin{aligned} I_{net} &= I_{cation} - I_{anion} \\ &= \frac{\pi d\sigma}{L}\frac{k_B T}{\eta \lambda_B}\Delta\log(C_{EH\_total}/C_{EL\_total}) + \pi(\frac{d}{2})^2 FD_{cation}\frac{\Delta C_{E\_cation}}{L} \\ &\quad -\frac{1}{2}\pi[(\frac{d}{2}-\lambda_{DH})^2 + (\frac{d}{2}-\lambda_{DL})^2]FD_{anion}\frac{\Delta C_{E\_anion}}{L} \end{aligned} \quad (16)$$

With the combination of both diffusion currents contributed by counterions and coions under the salt gradient, the theoretical expression Eq. 16 is derived for the net diffusion current $I_{net}$ inside the nanopore. Figure 8 illustrates the net diffusion current in nanopores of different lengths under different simulation conditions, which can be predicted by Eq. 16.

In the cases with ultra-short nanopores or under high salt gradients, significant concentration polarization[27] and diffusio-osmotic flow[44] may appear across and inside the nanopores, respectively. These phenomena may cause instability[45, 46] in the theoretical prediction of diffusion current with the derived equations, which requires more extensive and in-depth analysis.

Due to the effects of the high confinement and surface charges, ionic diffusion inside charged nanopores is a complicated process. Besides the pore geometry, surface charges, and solution conditions considered in this work, ionic diffusion can also be affected by the fluid flow inside the nanopore because directional ionic diffusion can induce diffusio-osmotic flow,[23, 47] which can be modulated with many factors at the solid-liquid interfaces,[48-50] such as roughness,[51, 52] slip length,[20, 53, 54] hydrophobicity,[55, 56] and hydrophilicity,[57, 58] by the presence of viscous friction, which may lead to an influence on the ion diffusion process. Currently, our model does not include the effects of hydrodynamic slip associated with flow through the nanopore. While discussions of interfacial hydrodynamics slip for fluid flow through nanopores are of fundamental importance, and this aspect will be our focus in future research.

## IV. CONCLUSIONS

Diffusive ionic transport through charged nanopores under concentration gradients has been systematically investigated by simulations. In long nanopores, the ionic diffusion current $I$ and the pore length $L$ satisfy the reciprocal relationship $I \propto \alpha/L$. The value of $\alpha$ depends on the cross-sectional areas of the passageway of coions and counterions inside nanopores. For the diffusion current of coions, the coefficient $\alpha$ has a quadratic relationship with the pore diameter. While $\alpha$ in the diffusion current of counterions can be described with the superposition of a quadratic relationship and a first power of the pore diameter. Because of concentration polarization across the nanopore, the diffusion current in neutral short nanopores can be predicted with the effective concentration difference, instead of the nominal one. Based on our simulations, equations are derived to describe the effective concentration differences across nanopores which can provide accurate theoretical predictions under various conditions. Taking into account the respective transport passageways of anions and cations, as well as the effective ion concentrations at both pore orifices, the theoretical prediction of the diffusion currents through nanopores is proposed, which are well suitable for nanopores longer than 100 nm and without EDLs overlapping. Our results may provide a better understanding of the microscopic details in ionic diffusion under confinement, and quantitative theoretical prediction of the induced ionic diffusive current through nanopores.

## SUPPLEMENTARY MATERIAL

See supplementary material for simulation details and additional simulation results.

## ACKNOWLEDGMENTS

The authors thank the support from the National Natural Science Foundation of China (Grant No. 52105579), the Basic and Applied Basic Research Foundation of Guangdong Province (2023A1515012931), the Natural Science Foundation of Shandong Province (ZR2020QE188), the Natural Science Foundation of Jiangsu Province (BK20200234), and the Qilu Talented Young Scholar Program of Shandong University.

## AUTHOR DECLARATIONS

**Conflict of Interest**

The author has no conflicts to disclose.

**Author Contributions**

Long Ma: Methodology (equal); Formal analysis; Software (lead); Validation; Writing – original draft (lead); Writing – review & editing (equal). Zihao Gao: Software (equal); Data curation; Visualization. Jia Man: Data curation (equal); Supervision (equal). Jianyong Li: Supervision (equal); Investigation (equal). Guanghua Du: Supervision (equal). Yinghua Qiu: Conceptualization; Resources; Supervision; Methodology (lead); Investigation (lead); Writing – original draft (equal); Writing – review & editing (lead); Funding acquisition.

## DATA AVAILABILITY

The data that support the findings of this study are available from the corresponding author upon reasonable request.

**NOMENCLATURE**

| | |
|---|---|
| $\varphi$ | Electrical potential, V |
| $R$ | Gas constant, J kg$^{-1}$ K$^{-1}$ |
| $T$ | Temperature, K |
| $F$ | Faraday's constant, C mol$^{-1}$ |
| $p$ | Pressure, Pa |
| $\varepsilon$ | Dielectric constant, F m$^{-1}$ |
| $\mu$ | Viscosity of solution, Pa s |
| **u** | Velocity of solution, m/s |
| $\nabla$ | gradient operator |
| $i$ | Cation or anion |
| $z_i$ | Valence of species $i$ |
| **J**$_i$ | ionic flux of species $i$ |
| $C_i$ | Concentration of species $I$, mol L$^{-1}$ |
| $D_i$ | Diffusion coefficient of species $i$, m$^2$ s$^{-1}$ |
| $D_{anion}$ | Diffusion coefficient of anions, m$^2$ s$^{-1}$ |
| $D_{cation}$ | Diffusion coefficient of cations, m$^2$ s$^{-1}$ |
| $N$ | Number of ion species in aqueous solution |
| $\sigma$ | Surface charged density, C m$^{-2}$ |
| **n** | Unit normal vector |
| $k_B$ | Boltzmann constant, J K$^{-1}$ |
| $\eta$ | Water viscosity, Pa s |
| $S$ | Reservoir boundary |
| $\alpha$ | Coefficient of the fitting equation $\alpha/L$ |
| $L$ | Pore length, nm |
| $d$ | Pore diameter, nm |
| $\lambda_D$ | Thickness of the EDLs, nm |
| $\lambda_{DH}$ | Thickness of the EDLs on the high-concentration side, nm |
| $\lambda_{DL}$ | Thickness of the EDLs on the low-concentration side, nm |
| $\lambda_B$ | Bjerrum length, nm |
| $C_H$ | High salt concentration, mol L$^{-1}$ |
| $C_L$ | Low salt concentration, mol L$^{-1}$ |
| $S_{EDLs}$ | Surface area of the EDLs region, nm$^2$ |
| $S_{center}$ | Cross-sectional area of the central region beyond the EDLs region, nm$^2$ |
| $\Delta C$ | Concentration difference between the high- and low-concentration sides, mol L$^{-1}$ |
| $\Delta C_{E\_NCS}$ | Effective concentration difference between the high- and low-concentration sides in the neutral nanopore, mol L$^{-1}$ |
| $\Delta C_{E\_anion}$ | Effective concentration difference of anions across the nanopore, mol L$^{-1}$ |
| $C_{EH\_anion}$ | Effective concentration of anions at the high-concentration side, mol L$^{-1}$ |
| $C_{EL\_anion}$ | Effective concentration of anions at the low-concentration side, mol L$^{-1}$ |
| $\Delta C_{E\_cation}$ | Effective concentration difference of cations across the nanopore, mol L$^{-1}$ |
| $C_{EH\_cation}$ | Effective concentration of cations at the high-concentration side, mol L$^{-1}$ |
| $C_{EL\_cation}$ | Effective concentration of cations at the low-concentration side, mol L$^{-1}$ |
| $C_{EH\_total}$ | Total effective concentration at the orifice of the high-concentration side, mol L$^{-1}$ |
| $C_{EL\_total}$ | Total effective concentrations at the orifice of the low-concentration side, mol L$^{-1}$ |
| $I$ | Diffusion current, nA |
| $I_{NCS}$ | Diffusion currents of cations and anions in neutral nanopore, nA |
| $I_i$ | Diffusion current of ions, nA |
| $I_{anion}$ | Diffusion current of anions, nA |
| $I_{cation}$ | Diffusion current of cations, nA |
| $I_{net}$ | Net diffusion current, nA |